\RequirePackage{filecontents}

\RequirePackage{snapshot}
\documentclass[aps,prl,reprint,showpacs,secnumarabic,superscriptaddress]{preprint}

\renewcommand\documentclass[2][]{}
\let\origparagraph=\paragraph
\AtBeginDocument{%
  \let\paragraph=\origparagraph
}
\makeatother
\documentclass[aps,prl,reprint,showpacs,floatfix]{revtex4-1}

\usepackage[T1]{fontenc}
\usepackage{amsmath}
\allowdisplaybreaks
\usepackage{textcomp}
\usepackage{txfonts}
\usepackage{tgtermes}
\renewcommand*{\vec}[1]{\boldsymbol{#1}}
\usepackage{microtype}
\usepackage{graphicx}
\graphicspath{{figures/}}
\usepackage[usenames,dvipsnames]{color}
\usepackage[breaklinks=true,
pdfauthor={Meng Wen, Heiko Bauke, and Christoph H. Keitel},
pdftitle={Dynamical spin effects in ultra-relativistic laser pulses}]{hyperref}

\makeatletter
\def\paragraph#1{%
  \@startsection%
    {paragraph}%
    {4}%
    {\parindent}%
    {\z@}%
    {-1sp}%
    {\normalfont\normalsize\itshape}{#1.---}%
}%
\makeatother

\newcommand*{\D}{\mathrm{d}}

\begin{document}

\title{Dynamical spin effects in ultra-relativistic laser pulses}

\author{Meng Wen}
\email{meng.wen@mpi-hd.mpg.de}
\affiliation{Max-Planck-Institut f{\"u}r Kernphysik, Saupfercheckweg~1,
  69117 Heidelberg, Germany} 

\author{Heiko Bauke}%
\email{heiko.bauke@mpi-hd.mpg.de}
\affiliation{Max-Planck-Institut f{\"u}r Kernphysik, Saupfercheckweg~1,
  69117 Heidelberg, Germany} 

\author{Christoph H. Keitel}%
\affiliation{Max-Planck-Institut f{\"u}r Kernphysik, Saupfercheckweg~1,
  69117 Heidelberg, Germany} 

\date{\today}

\begin{abstract}
  The dynamics of single laser-driven electrons and many particle
  systems with spin are investigated on the basis of a classical theory.
  We demonstrate that the spin forces can alter the electron dynamics in
  an ultra-relativistic laser field due to the coupling of the
  electron's spin degree of freedom to its kinematic momentum.
  High-energy electrons can acquire significant spin-dependent
  transverse momenta while passing through a counterpropagating
  ultra-relativistic infrared laser pulse.  Numerical calculations show
  that the deflection of the electrons by the laser pulse is determined
  by the laser intensity, the pulse duration, and the initial spin
  orientation of the electron.  We complement our investigation of these
  dynamical spin effects by performing particle-in-cell simulations and
  point out possibilities of an experimental realization of the
  predicted effect with available laser parameters.
\end{abstract}


\maketitle

\paragraph{Introduction}

The spin was introduced as an intrinsic property of the electron in
order to explain the emission spectra of alkali metals and the
Stern-Gerlach experiment. Spin dynamics and spin effects were widely
investigated, e.\,g., in semiconductors
\cite{Li:etal:2013:Femtosecond_switching}, diamond
\cite{Blok:Bonato:Markham:Twitchen:Dobrovitski:Hanson:2014:Nature_Physics},
graphene~\cite{
  Guttinger:2010:Spin_States}, quantum plasmas
\cite{Marklund:Brodin:2007:Dynamics_of}, gases
\cite{Beeler:etal:2013:spin_Hall_effect}, and also undulators
\cite{Mane:2012:Radiative_spin-flip}. It appears naturally in the
framework of relativistic quantum mechanics governed by the Dirac
equation.  However, a classical description of the electron spin may be
found phenomenologically or via a correspondence principle
\cite{Silenko:2008:Foldy-Wouthyusen_transformation,Walser:Szymanowski:Keitel:1999:EPL}.
The classical theory of spin was first laid down by Frenkel
\cite{Frenkel:1926:Elektrodynamik_des} and Thomas
\cite{Thomas:1926:Motion_of,Thomas:1927:kinematics} and further
developed by Bargmann, Michel, and Telegdi (BMT)
\cite{Bargmann:Telegdi:1959:Precession_of} and others. It is commonly
used to study the spin precession, e.\,g., in gravitation
\cite{PhysRevD.88.064037}, in ferromagnetic crystals
\cite{SPIN_RELAXATION,PhysRevLett.110.147201}, and in high-energy
physics
\cite{Mane:Shatunov:etal:2005:Spin-polarized_charged,Vieira:Huang:etal:2011:Polarized_beam}.

The investigation of spin effects has become relevant in particular due
to recent developments in particle accelerators and the availability of
high-intensity lasers \cite{Mourou:Tajima:etal:2006:Optics_in}.
Electron sources that utilize laser-driven acceleration provide electron
bunches with the size of a few micrometers
\cite{Lundh:etal:2011:Few_femtosecond}, a divergence of a few
milliradians \cite{PhysRevLett.109.064802}, an energy of hundreds of
megaelectronvolts and an energy spread of 1\,\%
\cite{PhysRevLett.102.164801}.  Laser intensities have reached the
ultra-relativistic domain at $10^{22}\,\text{W/cm}^2$
\cite{Yanovsky:Chvykov:etal:2008:Ultra-high_intensity-} and push
research in laser-matter interaction into the quantum electrodynamics
domain
\cite{Mourou:etal:2007:Relativistic_laser-matter_interaction,Di-Piazza:Muller:etal:2012:Extremely_high-intensity}.
Predicted phenomena such as non-dipole effects
\cite{Klaiber:etal:2013:Under-the-Barrier}, radiation reaction
\cite{PhysRevLett.102.254802,Hadad:Labun:etal:2010:Effects_of,Harvey:Heinzl:etal:2011:Symmetry_breaking,Ji:Pukhov:Kostyukov:Shen:Akli:2014:PRL,Green:Harvey:2014:PRL},
vacuum-polarization effects
\cite{Di-Piazza:Hatsagortsyan:etal:2008:Nonperturbative_Vacuum-Polarization},
and pair production \cite{Schwinger:1951:On_Gauge} may be realized
experimentally.  In particular, spin effects are expected to occur at
ultra-high intensities
\cite{Walser:Urbach:etal:2002:Spin_and,PhysRevLett.93.053002,Brodin:Marklund:etal:2011:Spin_and,Ahrens:etal:2012:spin_dynamics,Boca:Dinu:etal:2012:Spin_effects,PhysRevA.86.022109,Klaiber:etal:2014:Spin_dynamics}.
Angular-momentum related properties of electron bunches, such as
vortices in plasmas \cite{PhysRevLett.112.115002}, beam angular-momentum
\cite{PhysRevLett.111.135002}, and spin
polarization~\cite{Phys.Plasmas.21.023104.2014,Aulenbacher:EurPhysJST_198_361_2011}
have been studied.  To describe spin effects of energetic particles in
fields
\cite{Huang:Kim:2007:PhysRevSTAB,Esarey:Schroeder:Leemans:2009:RevModPhys,Corde:Ta_Phuoc:Lambert:Fitour:Malka:Rousse:Beck:Lefebvre:2013:RevModPhys}
and especially in the complex environment of plasmas we have to rely on
classical physics.  This can be accomplished by considering additional
terms in the classical equation of motion of charged particles, as it is
realized for QED effects in the Landau-Lifshitz equation
\cite{Landau1980Classical,Tamburini2011181,PhysRevE.84.056605,Tamburini:Keitel:etal:2014:Electron_dynamics}.

In this letter we study spin dynamics of free electrons classically in
ultra-relativistic laser fields.  We demonstrate that the coupling
between spin and kinematic momentum can modify the electron's spatial
motion significantly for feasible relativistic parameters as compared to
spinless particles.

\paragraph{Equations of motion for electrons with spin}

In classical theories, the spin of an electron with rest mass $m$ and
charge $q=-e$ is characterized by a unit vector $\vec{s}$, corresponding
to the direction of the spin's quantum mechanical expectation value.
The magnetic moment $\vec \mu$ of an electron is proportional to the
spin vector $\vec{\mu}=g\mu_\mathrm{B}\vec{s}/2$, with the gyromagnetic
factor $g\approx 2$, the Bohr magneton $\mu_\mathrm{B}={q\hbar}/({2mc})$
and $c$ and $\hbar$ denoting the speed of light and the reduced Planck
constant, respectively.  Following Frenkel
\cite{Frenkel:1926:Elektrodynamik_des}, it is convenient to incorporate
the spin dynamics into the electron's equations of motion by introducing
the spin tensor $\Pi_{\alpha\beta}$, which is defied as
\cite{Classical_spin_theory}
\begin{equation}
  \Pi_{\alpha\beta} =   
  \begin{pmatrix}
    0 & -T_x & -T_y & -T_z  \\
    T_x &    0 & -\Pi_z &  \Pi_y  \\
    T_y &  \Pi_z &    0 & -\Pi_x  \\
    T_z & -\Pi_y &  \Pi_x &    0 
  \end{pmatrix} =: (\vec{T},\vec{\Pi})\,,
\end{equation}
with $\vec T$ denoting the electric moment %
$ \vec{T}= \gamma \vec{\beta}\times \vec{s}$ and $\vec \Pi$ the magnetic
moment $ \vec{\Pi}= \gamma \vec{s}
-{\gamma^2}(\vec{\beta}\cdot\vec{s})\vec{\beta}/({\gamma+1})$.  Here
$\gamma$ is the Lorentz factor and $\vec{\beta}={\vec{v}}/{c}$ the
normalized velocity of the electron. Then the spin potential energy is
induced by the electron's magnetic moment $-\vec{\mu} \cdot
\vec{B}'=-{g\mu_\mathrm{B}} \Pi_{\alpha\beta} F^{\alpha\beta}/{4}$, in
which %
$\vec{B}'=\gamma(\vec{B}+\vec{E} \times \vec{\beta}) -
\mbox{${\gamma^2}\vec{\beta}(\vec{\beta}\cdot\vec{B})/({\gamma+1})$}$
represents the magnetic field in the electron's rest frame and %
$F^{\alpha \beta}=\partial^\alpha A^\beta - \partial^\beta A^\alpha
=:(\vec{E},\vec{B}) $ indicates the four-tensor of the electromagnetic,
the corresponding four-vector potential $A^\alpha$, and the electric
$\vec E$ and magnetic $\vec B$ fields in the laboratory frame
\cite{Jackson:1998:Classical_Electrodynamics}.  Requiring $\gamma
\mathcal{L}$ to be Lorentz invariant
\cite{Walser:Keitel:2001:Letters_in_Mathematical_Physics}, the
Lagrangian function $\mathcal{L}$ of the relativistic particle is given
by \cite{Barut:1980:Electrodynamics}
\begin{equation}
  \label{L}
  \mathcal{L} =
  -\frac{mc^2}{\gamma} + 
  \frac{g\mu_B}{4\gamma}\Pi_{\alpha\beta}F^{\alpha\beta}-
  \frac{q}{c}v_\alpha A^\alpha,
\end{equation}
where $v_{\alpha}=(c,-\vec{v})$ is the four-velocity of the particle and
$qv_\alpha A^\alpha/c$ equals the electromagnetic potential energy.
The electron's equations of motion for the position $\vec r$ and the
momentum $\vec p=m\gamma\vec v$ follow from the Lagrangian \eqref{L} via
the Euler-Lagrange equations as
\begin{subequations}
  \label{euler-lagrange}
  \begin{align}
    \label{dpdt}
    \eta \frac{\mathrm{d}}{\mathrm{d}t}\vec{p} & =  
    q(\vec{E}+\vec{\beta}\times\vec{B}) + \vec{f}_{S} \,,\\
    \label{drdt}
    \frac{\mathrm{d}}{\mathrm{d}t}\vec{r} & = \frac{\vec p}{m\gamma}\,,
    \intertext{with}
    \label{spinforce}
    \vec{f}_{S} & = 
    \frac{g}{2} \gamma\mu_\mathrm{B} \left(
      \frac{\vec\nabla}{\gamma^2} + 
      \vec{\beta} \frac{\mathrm{d}}{c \mathrm{d} t} 
    \right)
    ( \vec{s}\cdot \vec{B}' ) \,,\\
    \eta & = 
    1- \frac{g}{2}\frac{\mu_{\mathrm{B}}}{mc^2}\vec{s}\cdot \vec{B}'\,,
  \end{align}
\end{subequations}
in agreement with Ref. \cite{Barut:1980:Electrodynamics}.  The force in
\eqref{dpdt} acting on the electron is given by the well-known Lorentz
force plus the spin force $\vec{f}_{S}$ that both are modified by the
spin potential related factor $\eta$.  Note that $\D \vec B'/\D t$ can
also be written as $-\vec\nabla \times\vec E' + \mbox{$(c\vec \beta
  \cdot\vec\nabla)\vec B'$}$ by introducing the electric field $ \vec
E'$ in the electron's rest frame that is related to $\vec B'$ via the
Maxwell-Faraday equation. Thus, the spin force \eqref{spinforce}
originates from the inhomogeneities of electromagnetic fields.
Furthermore, the electron's momentum couples to the spin that precesses
in the electromagnetic fields under the effect of the BMT equations
\cite{Bargmann:Telegdi:1959:Precession_of}.  Taking into account the
electron's equation of motion \eqref{euler-lagrange} the BMT equations
are given by
\begin{subequations}
  \label{dsdt}
  \begin{align}
    \frac{\mathrm{d}\vec{s}}{\mathrm{d}t} & =
    \vec{\Omega}_{0}+\vec{\Omega}_{1}\,,\\
    \vec{\Omega}_{0} & = 
    \frac{q}{mc} \vec{s}\times \left[
      {\eta_1}\vec{B} - 
      {\eta_2}\vec{E} \times \vec{\beta} -
      \eta_3\frac{\gamma}{\gamma+1}\vec{\beta}\left(\vec{\beta}\cdot\vec{B}\right)
    \right]\,,\\
    \vec{\Omega}_{1} & = 
    \frac{g}{2}\frac{\mu_{\mathrm{B}}}{\eta mc} \frac{1}{\gamma+1}\vec{s}\times
    \left(\vec{\beta}\times\vec\nabla\right)\left(\vec{s}\cdot \vec{B}'\right)\,,
  \end{align}
\end{subequations}
where $\vec{\Omega}_{0}$ is the contribution to the spin precession by
the electromagnetic fields in zeroth order with
$\eta_1=g/2-\mbox{$(\gamma-1)$}/(\gamma\eta)$,
$\eta_2=g/2-\gamma/\eta/(\gamma+1)$, and $\eta_3=g/2-1/\eta$; while
$\vec{\Omega}_{1}$ accounts for field gradients
\cite{Good:1962:PhysRev}.

Radiation reaction may cause additional damping forces that may be
incorporated into \eqref{dpdt} via the additional force term $\vec f_R$,
which is part of the four-vector $f_R^\alpha = (f_R^0, \vec f_R) = r_e
(q \gamma \partial_\gamma F^{\alpha\beta}v_\beta v^\gamma + q^2
F^{\alpha\beta}F_{\beta\gamma}v^\gamma / mc - q^2 \gamma^2
F^{\beta\gamma} F_{\gamma\delta} v^\delta v_\beta v^\alpha/mc^3)$ with
$r_e\approx 1.18 \times 10^{-8} \,\mbox{\textmu m}$
\cite{Teitelboim:Nuovo_Cimento:1980}.  In all calculations of this work
we take into account the effects of radiation damping by adding $\vec
f_R$ on the right hand side of \eqref{dpdt}.  Modifications of the BMT
equations \eqref{dsdt} due to radiation damping are found negligible for
the parameters employed here.

\paragraph{Electrons counterpropagating laser fields}

We consider a plane wave with circular polarization propagating along
the $x$~axis with its vector potential \mbox{$\vec A =\vec a m c^2/e$},
and electric and magnetic fields \mbox{$\vec E=-\partial_t \vec A /c$}
and \mbox{$\vec B = \vec \nabla \times \vec A$}, where the scaled vector
potential $\vec a = (0,a_y,a_z)$ with \mbox{$a_y + ia_z =a_0 \,
  \mathrm{sin}^2(\pi\tau/T) \mathrm{exp}(i\omega_L \tau)$} and the
scaled field's amplitude $a_0=(I_0/\pi)^{1/2}e\lambda_L/(mc^{5/2})$
depends on the laser intensity $I_0$, the phase
$\omega_L\tau=\omega_L(t-x/c)$, the duration $T$ of the laser pulse, and
the angular frequency $\omega_L=2\pi/\tau_L=2\pi c/\lambda_L$ with the
laser's wavelength $\lambda_L$ and its period $\tau_L$. In the case of
interest here with $\gamma_0 \gg 1$, the $\vec \nabla / \gamma^2$ term
in~\eqref{spinforce} can be neglected, where $\gamma_0$ is the Lorentz
factor at time $t=0$. Then the dominant term of the spin force
originates from the total time derivative of the magnetic field, which
includes a partial time derivative and spatial gradients
\mbox{$\mathrm{d}\vec B'/\mathrm{d}t=(\partial/\partial t + c\vec \beta
  \cdot \vec\nabla)\vec B'$}.

\begin{figure}[t]
  \includegraphics[width=0.95\linewidth]{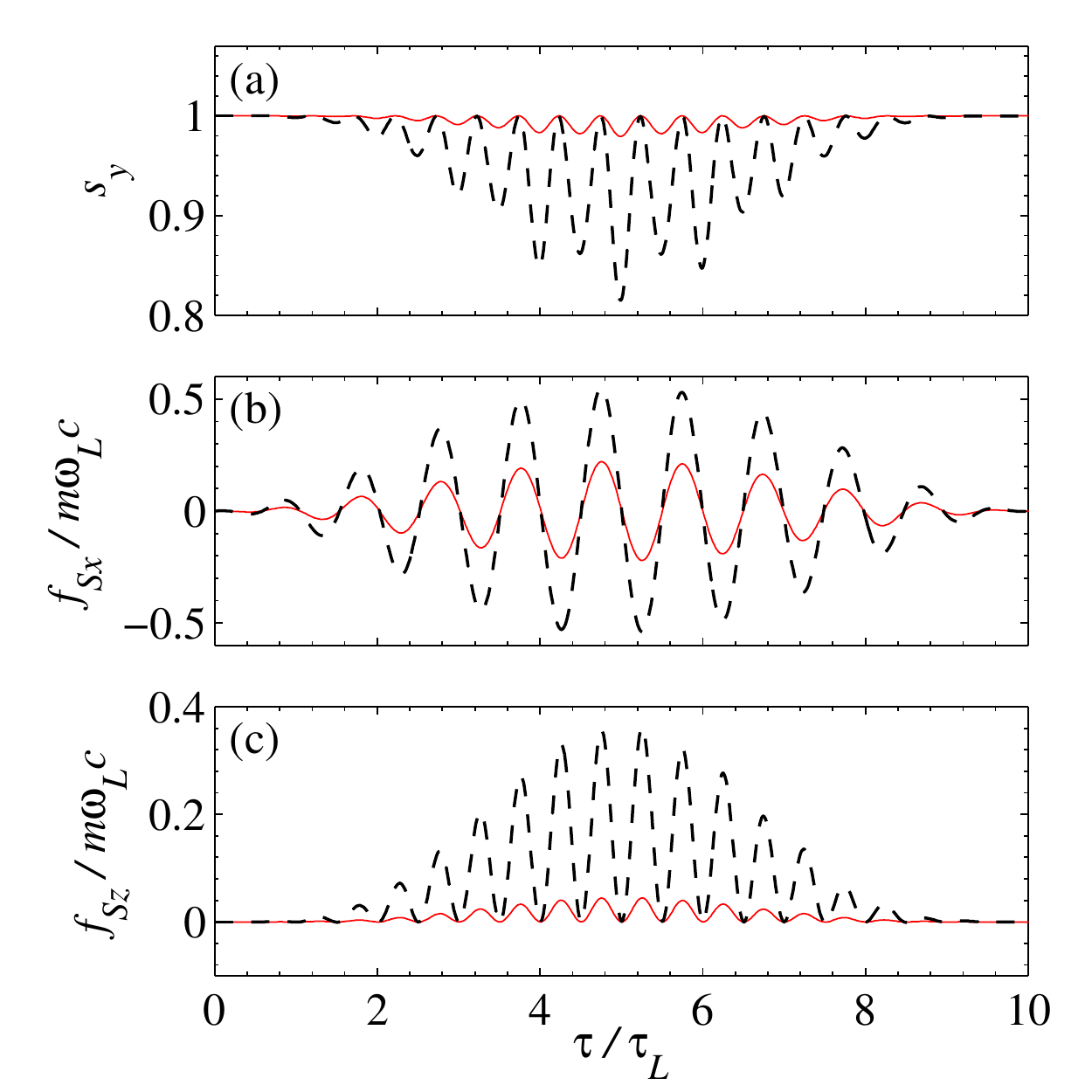}
  \caption{(color online) (a) The $y$~component of the spin vector as it
    evolves under Eq.~\eqref{dsdt} as a function of
    $\tau=t-x(t)/c$. (b), (c) The evolution of the $x$ and
    $z$~components of the spin force $\vec{f}_{S}$ for two different
    peak intensities.  Solid (red) and dashed (black) curves correspond
    to an intensity of $I_0=4.28\times10^{20} \,\mathrm{W/cm^2}$
    ($a_0=10$) and $I_0=3.85\times10^{21}\,\mathrm{W/cm^2}$ ($a_0=30$),
    respectively.  The laser's wavelength equals
    $\lambda_L=800\,\mathrm{nm}$ and the pulse duration is $T=10\tau_L$
    with $\tau_L=\lambda_L/c$. The electron's initial energy corresponds
    to $\gamma_0=50$.}
  \label{fs}
\end{figure}

The magnitude of the spin force can be estimated from the dominant term
of Eq.~\eqref{spinforce}, i.\,e., $\vec f_S \approx g \mu_\mathrm{B}
\vec p / (2mc^2) \, \mathrm{d} (\vec s \cdot \vec B')/\mathrm{d}t$,
which depends on the particle's momentum $\vec p$ as well as on the spin
vector $\vec s$ and $\vec B'$. The momentum of an energetic electron
counterpropagating in a plane wave with $\gamma_0 > a_0$ remains in the
longitudinal direction $p_x \approx -\gamma mc$ and the electromagnetic
fields cause $\vec{p}_\perp \approx\vec a mc$.  Thus, the velocity
$\vec\beta$ is almost perpendicular to $\vec B$ and therefore $\vec B'
\approx 2\gamma \vec B$.  Furthermore, the spin vector $\vec s$
oscillates in the laser field but remains close to its initial value,
especially for relatively low laser intensities.  This is demonstrated
in Fig.~\ref{fs}(a), which shows the $y$~component of the spin vector
for an electron with initial spin direction $\vec{s}=\vec{e}_y$.  Thus,
with $\vec s \approx\vec e_y$ and $\vec B' \approx 2\gamma \vec B$, we
get
\begin{math}
  \D (\vec s \cdot \vec B')/\D t \approx
  \mbox{$2\gamma(1-\beta_x) a_z m\omega_L^2 c/e $}
\end{math} and the counterpropagating setup, where $(1-\beta_x)\approx
2$, maximizes the spin force. Finally, the dominant term of the spin
force can be expressed by the laser field $\vec a$ and the electron's
energy $\gamma$ as $\vec f_S\approx( - \gamma, a_y, a_z)\,a_z\gamma g m
\omega_L c\lambda_e/\lambda_L$ with the Compton wavelength
$\lambda_e=4\pi\mu_\mathrm{B}/q$ ($\lambda_e/\lambda_L= 3.03\times
10^{-6}$ for a laser wavelength of $\lambda_L=800\,\mathrm{nm}$).  Thus,
$\vec{f}_{S}$ increases with rising laser intensity and initial electron
energy.  For relativistic particles in strong fields it may become even
comparable to the Lorentz force. The relative weight of the spin force
$\vec{f}_{S}$ in \eqref{dpdt} reaches around 6\,\% of the amplitude of
the Lorentz force for $a_0\sim\gamma_0\sim10^2$.

When a spinless particle passes through a time-symmetric laser pulse,
its transverse momentum does not increase due to the symmetric
oscillations of the directions of the Lorentz force
\cite{LAWSON:1979:LASERS_AND_ACCELERATORS} and the radiation reaction
force \cite{Piazza:Lett_Math_Phys:2008}.  This symmetry is broken by the
spin force $\vec f_{S}$ if the particle features a spin degree of
freedom, e.\,g., for electrons. The spin force $\vec{f}_{S}$ oscillates
in the laser field, see Figs.~\ref{fs}(b) and~(c). The $x$~component of
the spin force \mbox{$f_{Sx} \propto \gamma^2 a_z$} oscillates
synchronously with the magnetic field and its effect averages out over
several laser cycles, see Fig.~\ref{fs}(b).  Similarly, the $y$
component also oscillates symmetrically $f_{Sy}\propto \gamma a_y a_z$
with vanishing net effect.  However, the $z$~component of the spin
force, which is in leading order \mbox{$f_{Sz} \propto \gamma a_z^2$},
varies with the frequency $2\omega_L$ and remains positive for all
times, see Fig.~\ref{fs}(c). Consequently, a spin-force effect is
accumulated to a significant extra momentum over the duration of the
laser pulse.  
Note that the contribution of laser fields to $f_{Sz}$ arise only via
$a_z$. Thus, a linearly polarized pulse $\vec a = (0, 0, a_z)$
counterpropagating to an electron with initial spin $\vec s=\vec e_y$
would yield similar effects as circularly polarized pulses.

\begin{figure}
  \includegraphics[width=0.95\linewidth]{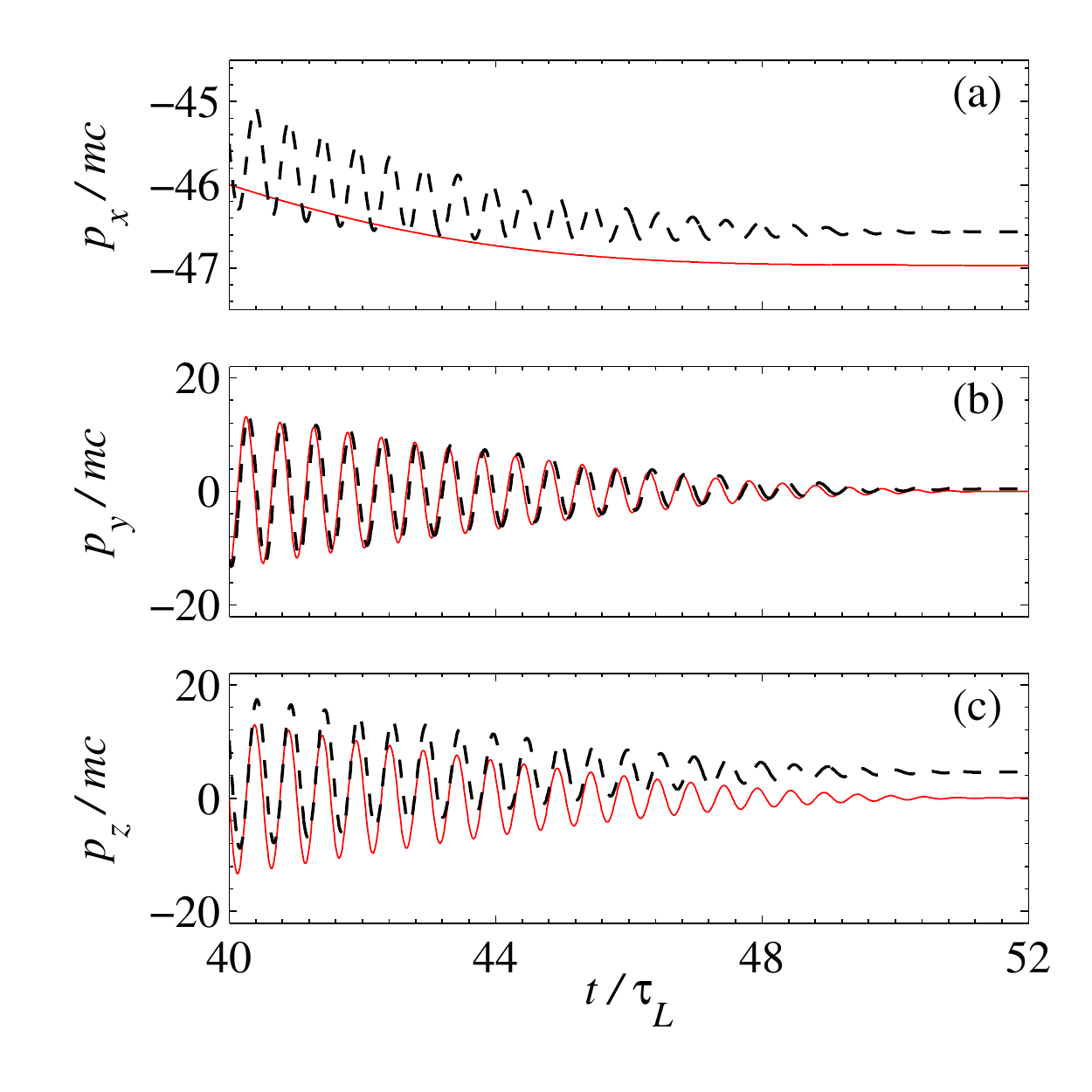}
  \caption{(color online) Evolution of the momenta $p_x$ (a), $p_y$ (b)
    and $p_z$ (c) at the end of the interaction with the laser pulse for
    particles with initially $\gamma_0=50$.  Solid (red) and dashed
    (black) curves correspond to spinless particles and electrons with
    spin, respectively.  The laser intensity is the same as in
    Fig.~\ref{fs}(b), with the duration $T=100\tau_L$.  }
  \label{momentum}
\end{figure}

The effect of the spin forces becomes explicit by comparing the
time-evolution of the momentum of spinless particles with that of
electrons as shown in Fig.~\ref{momentum}.  Initially, the particles
move along the $x$ coordinate to the left such that they encounter the
laser pulse at $t=0$ and $x=0$.  The equation of motion for spinless
particles involves the Lorentz force and radiation reaction yielding the
solid (red) curves in Fig.~\ref{momentum}.  In this case, the transverse
momenta vary symmetrically over the interaction time.  As a net effect
of radiation reaction the final longitudinal momentum $|p_x|=47 $ is
smaller than the initial $|p_{x0}|=50$. The dynamics changes, however,
when the spin is taken into account via the coupled
Eqs.~\eqref{euler-lagrange} and~\eqref{dsdt}.  The momentum evolution of
an electron in an ultra-relativistic laser pulse and with initial spin
$\vec s=\vec e_y$ is indicated by the dashed (black) curves in
Fig.~\ref{momentum}.  As a consequence of the spin force, the transverse
momentum components after interaction with the laser pulse differ from
the initial momenta.  Thus, the interaction of the electron with the
laser field has a non-vanishing net effect on the electron's transverse
momenta.  Furthermore, the damping of the longitudinal momentum is
stronger as compared to spinless particles via extra acceleration from
the spin force, see solid (red) and dashed (black) lines in
Fig.~\ref{momentum}.  For the parameters in Fig.~\ref{momentum}, for
example, the electron's initial momentum is $(-50, 0, 0)mc$ but its
final value yields $(-46.56, 0.44, 4.62)mc$. Further numerical
calculations indicate that if a focused laser pulse is considered
\cite{PhysRevLett.88.095005} the spin effect does not change
qualitatively as compared to the plane wave case as long as the focus
waist is larger than several laser wavelengths.

\begin{figure}[t]
  \centering
  \includegraphics[width=0.95\linewidth]{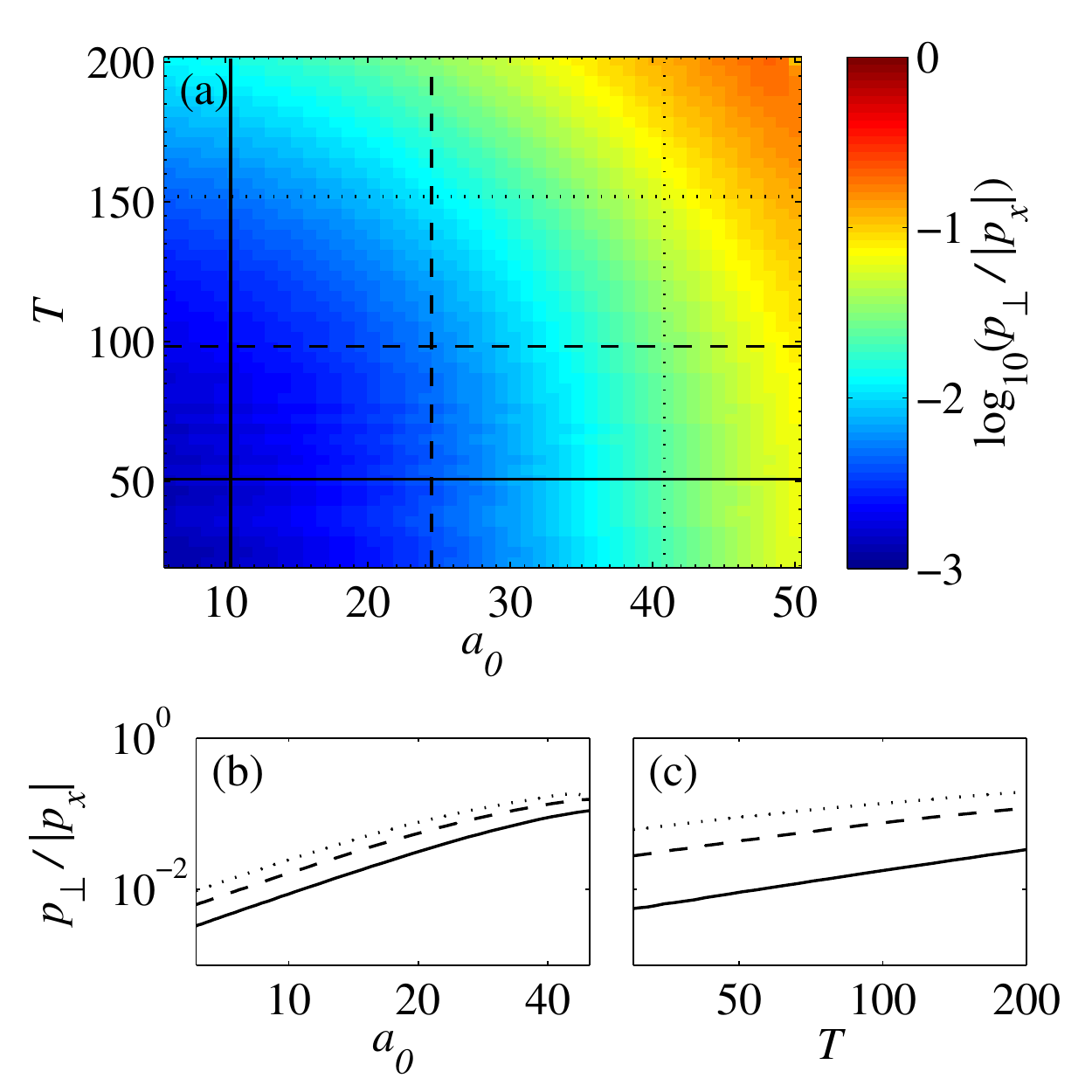}
  \caption{(color online) (a) The dependencies of the final deflection
    ${p_\perp}/{|p_x|}$ for an electron with initial spin $\vec s=\vec
    e_y$ on the laser field's amplitude $a_0$ and on the duration
    $T$. The solid, dashed, and dotted curves in (b) and (c) correspond
    to the horizontal and vertical curves in (a). The color in (a)
    refers to the logarithm of ${p_\perp}/{|p_x|}$.  The initial
    electron energy corresponds to $\gamma_0=50$. The quantum parameter
    $\chi_0 =4.4 \times 10^{-6} \gamma_0 a_0$
    \cite{Di-Piazza:Muller:etal:2012:Extremely_high-intensity} is $1.1
    \times 10^{-3} < \chi_0 < 1.1 \times 10^{-2}$ for the cases
    $5<a_0<50$ investigated here.}
  \label{depend}
\end{figure}

Experimentally the momentum transfer in the laser field may be verified
via the determination of the deflection ${p_\perp}/{|p_x|}$ with
$p_\perp =(p_y^2+p_z^2)^{1/2}$ instead of measuring the individual
momentum components.  For an unpolarized electron beam the deflection
will happen symmetrically around the laser's propagation axis.
Figure~\ref{depend} presents how the deflection of an electron that is
initially polarized along the $y$~direction depends on the laser's
amplitude $a_0$ and the pulse duration $T$.  The deflection does not
depend on $\gamma_0$ directly. However, $\gamma_0$ has to be at least
larger than $a_0/2$ to maintain the counterpropagating regime. The
deflection becomes larger for higher laser intensities and also growths
with the pulse duration $T$, see Fig.~\ref{depend}.  For the employed
field strengths $a_0$ and pulse durations $T$, the deflection
${p_\perp}/{|p_x|}$ depends quadratically on the field strength and
linearly on the pulse duration.  It can be estimated as $p_\perp/|p_x|
\approx (1/2)(\lambda_e/\lambda_L) a_0^2 (T/\tau_L)$.

\paragraph{Particle-in-cell simulations}

In order to investigate how the dynamical spin effect changes the
dynamics for a whole bunch of electrons we performed particle-in-cell
(PIC) simulations \cite{Wu:2011:JPIC} employing a 1D3V-model which is
one-dimensional in position and three-dimensional in velocity space.
The reduction to one dimension in position space is justified because
the electromagnetic field of the laser depends on the $x$~coordinate
only when laser focusing is negligible.  The PIC simulations include the
spin degree of freedom as an additional property of the pseudo particles
as well as radiation reaction.  In each time step the coupled equations
of motion Eqs.~\eqref{euler-lagrange} and~\eqref{dsdt} are integrated
for all pseudo particles. To simulate spin precession, a Boris'~rotation
\cite{Boris:1970:Relativistic_Plasma_Simulation,BirdsallLangdon} of the
spin is performed while its momentum rotates between the two stages of
half-accelerations in each time step.  The field gradients are
calculated via linear interpolation of the fields at adjacent cell
boundaries.

\begin{figure}
  \centering
  \includegraphics[width=0.95\linewidth]{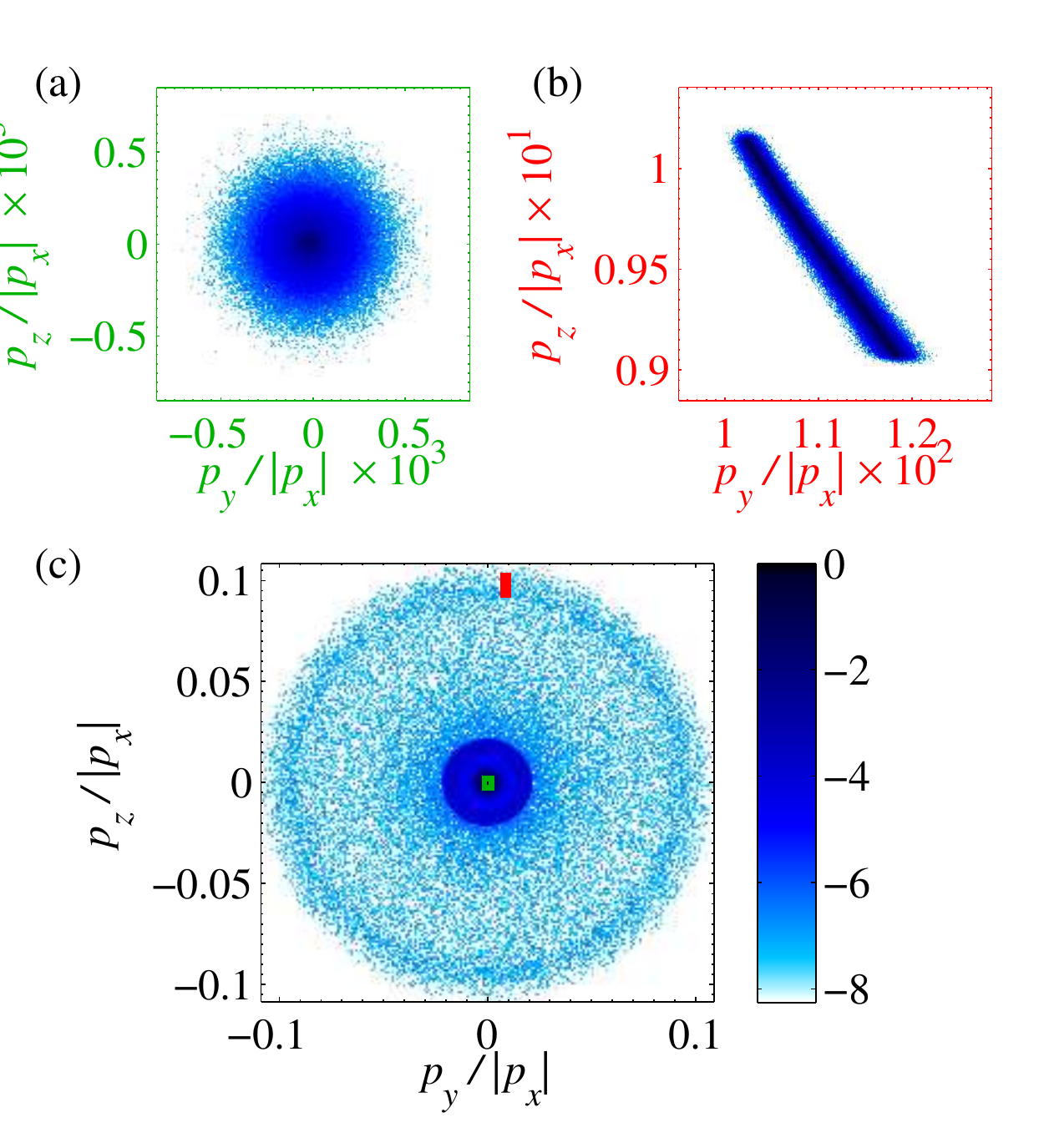}
  \caption{(color online) PIC simulated transverse momentum distribution
    of a particle bunch after interaction with a counterpropagating
    circularly polarized laser pulse for spinless particles (a), for
    initially $y$~polarized electrons (b), and for initially unpolarized
    electrons (c).  The color indicates the logarithm of the
    distribution's density in arbitrary units.  Note the different
    scales of the various subfigures.  The green and red rectangles in
    subfigure (c) indicate the regions of the momentum spaces that are
    shown in subfigures (a) and (b).  The laser pulse has an intensity
    of $I_0=3.85\times10^{21}\,\mathrm{W/cm}^2$ ($a_0=30$) and consists
    of 100 cycles.  Initially the bunch is directed strictly along the
    $x$~direction with an energy of each electron $\gamma_0=50$ and a
    divergence of 1\,mrad. The initial bunch has a length of
    $l_e=5\lambda_L=4\,\mbox{\textmu}\mathrm{m}$ and a density of
    $n_e=0.001n_c\approx1.7\times10^{18}\,\mathrm{cm^{-3}}$, where
    $n_c=\pi mc^2/(\lambda_L e)^2$ refers to the critical plasma
    density.}
  \label{thzthy}
\end{figure}

Figure~\ref{thzthy} shows the transverse momentum distribution of a
particle bunch after interaction with a counterpropagating circularly
polarized laser pulse for (a) spinless particles, (b) initially
$y$~polarized electrons, and (c) initially unpolarized electrons.  For
the spinless case shown in Fig.~\ref{thzthy}(a) the divergence of the
bunch changes marginally during its interaction with the laser pulse.
The electron-bunch's expansion by the Coulomb potential is very weak due
to the electrons' high energy and the low density.  If the spin of the
particles is taken into account, however, electrons of a fully polarized
bunch are deflected by the spin forces, see Fig.~\ref{thzthy}(b).  This
effect appears detectable experimentally.  The amount of the bunch's
deflection agrees with the prediction for the singe-electron case
considered in Fig.~\ref{depend}.  An unpolarized electron bunch with the
initial spin orientation distributed homogeneously in all directions is
considered in Fig.~\ref{thzthy}(c).  The final momentum distribution is
symmetric around the laser's propagation direction and most electrons
are deflected by an angle that is predicted by singe-electron
trajectories.  Thus, a significant increase of the divergence angle to
$\theta = 2\arctan(p_\perp/|p_x|) \approx 198\,\mathrm{mrad}$ results.

\paragraph{Discussion and conclusion}

We investigated electron motion in strong laser fields focussing on spin
effects.  For this purpose we applied classical equations of motion for
the electron's position, kinematic momentum, and its spin.  These
coupled equations were derived from a Lorentz invariant Lagrangian and
are a generalization of the well-known BMT equations.  We apply this
theory to the interaction of an electron bunch with a counterpropagating
high-intensity laser pulse.  In this setting, electrons are deflected
significantly by the spin force, which originates from the particles'
high energy, strong laser gradients as well as the spin component along
the magnetic field. Focusing and polarization states of the laser pulse
do not change this effect qualitatively. These results were obtained by
solving numerically the equations of motion for single electrons and
further confirmed by PIC simulations for an electron bunch with
parameters suitable for a possible experimental realization of this
effect.  In conclusion, the influence of the spin on the motion of an
electron should be taken into account in the indicated situations of
light-matter interaction at relativistic intensities with strong laser
inhomogeneities, long interaction time and highly preaccelerated
particles.

\begin{acknowledgments}
  We would like to thank Dr.~A.~Di~Piazza and Dr.~K.~Z.~Hatsagortsyan
  for valuable discussions.
\end{acknowledgments}

\bibliography{classical_spin_dynamics}

\end{document}